\documentclass[12pt]{article}
\input epsf
\usepackage{psfrag}
\usepackage{graphicx}
\newcommand {\eqref} [1] {(\ref {#1})}
\newcommand {\slsh} [1] {\not{\hbox{\kern-2pt${#1}$}}}

\newcommand {\be} {\begin{equation}}
\newcommand {\ee} {\end{equation}}
 \newcommand {\ber}{\begin{eqnarray*}}
 \newcommand {\eer} {\end{eqnarray*}}
\newcommand {\bea}{\begin{eqnarray}}
 \newcommand {\eea} {\end{eqnarray}}

\newcommand{\ba}{\begin{array}}
\newcommand{\ea}{\end{array}}

\def\dps{\displaystyle}
\newcommand{\half}{\frac{1}{2}}

\def\Acknowledgements{\bigskip  \bigskip {\begin{center} \begin{normalsize}
             \bf ACKNOWLEDGEMENTS \end{normalsize}\end{center}}}

\begin{document}
\noindent

\begin{titlepage}
\vskip 1cm
\rightline{hep-th/0410116}
\vskip 3cm
\centerline{{\huge Holography, Duality and Higher-Spin Theories}}
\vskip 2cm
\centerline{{ \large Anastasios C. Petkou}}
\vskip 0.3cm
\centerline{ \it University of Crete, Department of Physics}
\centerline{\it  71003 Heraklion, Greece}

\vskip 2cm

\begin{abstract}

I review recent work on the holographic relation  between higher-spin
theories in Anti-de Sitter spaces and conformal
field theories. I present the main results of studies concerning the
higher-spin holographic dual of the three-dimensional $O(N)$ vector
model. I discuss the special role played by  
certain double-trace deformations in Conformal Field Theories that have
higher-spin holographic duals. Using the canonical formulation I show
that duality
transformations in a $U(1)$ gauge theory on AdS$_4$ induce boundary
double-trace deformations. I argue that a similar effect takes 
place in the holography of 
linearized higher-spin theories on AdS$_4$.

\vspace{2cm}

{\it Based on a lecture given at the First Solvay Workshop,
  ``Higher-Spin Gauge Theories'', Brussels, May 12-14 2004.

Notes taken by K. Alkalaev.}
\vfill

\end{abstract}
\end{titlepage}

\section{Introduction}

Consistently interacting higher-spin (HS) gauge theories exist on Anti-de
Sitter spaces (see \cite{Vasiliev,Sagnotti} for recent reviews and
extensive literature on higher-spin gauge theories). It is then 
natural to ask whether HS theories have interesting holographic
duals. In this lecture I review recent work on the holographic
relation between HS theories and conformal field theories
(CFTs). After  some general remarks concerning the relevance of HS
theories to free CFTs, I present the main results in studies of the
three-dimensional critical $O(N)$ vector model that has  been suggested
to realize the holographic dual of a HS theory on AdS$_4$. Furthermore,
I discuss the special role the so-called double-trace
deformations seem to play in the dynamics of CFTs that have holographic
HS duals. I particular, I show that duality transformations of a
$U(1)$ gauge theory 
on AdS$_4$
 induce boundary double-trace deformations. I argue that a similar
 effect takes place in the implementation of holography to linearized
 HS theories on AdS$_4$.

\section{Higher-spin currents and the operator spectrum of  Conformal
  Field Theories}

The operator spectrum of $d$-dimensional Conformal Field Theories
(CFTs)  consists of an infinite
set of modules each containing one quasi-primary
operator  $\Phi(0)$ that is annihilated by the generator $\hat{K}_\mu$  of
special conformal transformations \cite{MS}
\be
\hat{K}_\mu \Phi(0) = 0\;,
\ee
as well as  an infinite number of descendants that are essentially the
derivatives of $\Phi(0)$.
Thus, quasi-primary operators carry irreps of the 
conformal group $SO(d,2)$ labeled  only by their spin $s$
and scaling dimension $\Delta$. As a consequence, the 2- and 3-pt
functions of quasi-primary operators are determined up to a number
of constant parameters (see for example \cite{OP}). 

Perhaps the most important property
of quasi-primary operators is that they form an
algebra under the (associative) operator product expansion (OPE)
\cite{Mack}. Given two such operators $A(x)$ and $B(x)$
we may expand their product in a (infinite) set of quasi-primary
operators $\{Q\}$ as
\be
A(x)B(0)  = \sum_{\{Q\}} C(x,\partial) Q(0)\;.
\ee
The
coefficients $C(x,\partial)$ are  fully determined in terms of the
spins, dimensions and 3-pt functions of the operators involved in the
OPE. Knowledge of the OPE is potentially sufficient to
determine all correlation functions of quasi-primary operators in CFTs.

In specific models all quasi-primary operators (with the exception of
the "singletons"), are composite
operators and determining their precise list, spins and dimensions is a hard
task which is done in practice by studying 4-pt functions.
For example, consider a scalar quasi-primary operator $\Phi(x)$ with
dimension $\Delta$. The OPE enables us to write 
\bea
\label{4pt}
\langle \Phi(x_1)\Phi(x_2)\Phi(x_3)\Phi(x_4)\rangle &=&\sum_{\{Q\}} C(x_{12},
\partial_2)C(x_{34},\partial_4)
\langle Q(x_2)Q(x_4)\rangle \nonumber \\
&=&= \frac{1}{(x_{12}^2x_{34}^2)^\Delta}
\sum_{\{\Delta_s,s\}}g_{\Delta_s,s} \,H_{\Delta_s,s} (v,Y)\nonumber \\
&=& \frac{1}{(x_{12}^2x_{34}^2)^\Delta}\sum_{\{\Delta_s,s\}}g_{\Delta_s,s}
\,v^{\frac{\Delta_s-s}{2}}Y^s\Bigl[1+O(v,Y)\Bigl] \;,
\eea
where we have used the standard harmonic ratios
\be
v =\frac{x_{12}^2x_{34}^2}{x_{13}^2x_{24}^2}\;,\qquad u =
\frac{x_{12}^2x_{34}^2}{x_{14}^2x_{23}^2}\;,\qquad Y
=1-\frac{v}{u},\qquad
x_{ij}^2 = |x_i-x_j|^2\;.
\ee
The coefficients $g_{\Delta_s,s}$ are the "couplings".
The expressions $H_{\Delta_s,s}(u,Y)$ are complicated but explicitly
known \cite{HPR} functions that give the contribution of an
operator with spin $s$ and dimension $\Delta$ to the OPE.  The terms 
$v^{(\Delta_s-s)/2}Y^s$  correspond to the leading  short-distance
behavior of the 4-pt function for $x_{12}^2\,,x_{34}^2\rightarrow 0$.

In the simplest case of a massless free CFT the set $\{Q\}$ includes,
among others,  all
symmetric traceless
quasi-primary operators with
spin $s$ and dimension $\Delta_s$. It is a general result in CFT that
if the above operators are also conserved, then their
dimensions are given by \footnote{The converse is not always true
  \cite{Ferrara}, therefore even if one finds terms such as (\ref{vY})
  in the OPE one should be careful in interpreting them.}
\be
\label{candim}
\Delta_s = d-2+s \,.
\ee
Therefore, whenever higher-spin conserved currents reside in the list of
quasi-primary operators each one contributes in (\ref{4pt}) a term of
the form
\be
\label{vY}
v^{\frac{d-2}{2}}Y^s\,.
\ee
Of course, irreps with canonical scaling dimension (\ref{candim})
are nothing but the massless UIRs of the $d+1$-dimensional   Anti-de
Sitter group.  

Suppose now that one is interested in the holographic
description of the 4-pt function (\ref{4pt}). For that, we notice that
from a classical action on Anti-de Sitter we get 
non-trivial boundary correlators by studying tree-level
bulk-to-boundary graphs \cite{Witten_adscft}. 
Hence, when the boundary 4-pt function contains terms like
(\ref{vY}) it is necessary to consider bulk massless currents.
On the other hand, 
the energy-momentum tensor always appears in the OPE of a 4-pt
function, it has dimension $d$ and spin-2. Its dimension remains
canonical. Its contribution to the 4-pt function is given by the
(rather complicated) function 
\bea
\label{EM_1}
H_{e.m.}(v,Y) &=& v F_1(Y) [1+O(v,Y)]\,,\\
\label{EM_2}
F_1(Y) &=& \frac{4Y^2-8Y}{Y^3} + \frac{4(-6+6Y - Y^2)}{Y^3}\ln(1-Y)
\longrightarrow Y^2 + ...
\eea
A concrete realization of the ideas above is provided by explicit
calculations
in  ${\cal N} =4$ SYM$_4$ via the AdS/CFT correspondence \cite{AFP}.
Consider the 4-pt function of the so-called lower dimension chiral
primary operators (CPOs)
of ${\cal N}=4$, which are scalar operators with dimension 2.  In the
free field theory limit we have
\bea
\label{free4pt}
&&\hspace{-2cm}\frac{\delta^{I_1I_2} \delta^{I_3I_4}}{400} \langle
Q^{I_1}(x_1) ... Q^{I_4}(x_4)  \rangle_{free}  =
\nonumber \\
&&= \frac{1}{(x^2_{12}x_{34}^2)^2}\Bigl( 1+\frac{1}{20}v^2
+\frac{1}{20}v^2(1-Y)^{-2}\nonumber \\
&&+\frac{4}{N^2}(\frac{1}{6}[v+v(1-Y)^{-1}]+
\frac{1}{60}v^2(1-Y)^{-1}) \Bigl)\nonumber \\
&&=\frac{1}{(x^2_{12}x_{34}^2)^2}\Bigl( ... +
\frac{4}{6N^2}\sum_{l=2}^{\infty}vY^l+...\Bigl)\;.
\eea
In the last line of (\ref{free4pt}) we see the contribution of
an infinite set of
higher-spin conserved currents in the {\it connected} part of the
correlator. The leading contribution comes from
the energy-momentum tensor. One can identify the contributions from
all the higher-spin currents and even calculate their "couplings",
after some hard work that involves the subtraction of the descendants of each
and every current. The
perturbative corrections to the connected part of the free result
above have the form
\be
[connected] = \frac{1}{N^2}[connected]_{free} + \frac{1}{N^2}
g^2_{YM}N\, F(v,Y)\;,
\ee
where
\be
F(v,Y) \sim \sum_l v\,Y^l \ln v + ...
\ee
The above terms can be attributed to an infinite set of "nearly
conserved" higher-spin currents i.e. quasi-primary
operators whose scaling dimensions have being shifted from their
canonical value as \cite{AFP}
\be
\Delta_{HS} \longrightarrow \Delta_{HS} +\gamma = 2+s+
(g^2_{YM}N)\eta_s+\cdots \;.
\ee
By the operator/state correspondence in CFTs, we may view the above
effect as a small deformation of the energy
spectrum of the theory. From the AdS side this deformation should
correspond to a
Higgs-like effect by which the initially massless higher-spin
currents acquire masses \cite{Porrati,Bianchi}.

In the context of the $AdS_5/CFT_4$ correspondence, one can calculate the
same 4-pt function using IIB supergravity \cite{AF}. The result is
highly non-trivial and looks like
\be
[connected]  = \frac{1}{N^2}\frac{1}{(x_{12}^2x_{34}^2)^2}\Big[v F_1(Y)
+O(v^2,Y)\Big]\,.
\ee
It is an astonishing fact that the expansion of such a non-trivial
function reveals the presence of {\it only} the energy momentum tensor
and the absence of {\it all} higher-spin currents. This shows how far
away supergravity is from a holographic description of perturbative
CFTs. This shows also the necessity to consider HS gauge theories
if we wish
to describe holographically
perturbative CFTs.

\section{Holography of the critical three-dimensional $O(N)$
  vector model }

A concrete proposal for the holographic correspondence between a CFT and
a HS gauge theory was made in \cite{KP}. It was there suggested that
the critical three-dimensional $O(N)$ vector model is the holographic
dual of the simplest HS gauge theory on AdS$_4$, a theory that
contains bosonic symmetric traceless even-rank tensors.
The elementary fields of the (Euclidean) three-dimensional $O(N)$
vector model are the scalars
\be
\Phi^a(x) \,,\,\,\,\,\, a=1,2,..,N\,,
\ee
constrained by
\be
\Phi^a(x)\Phi^a(x) =\frac{1}{g}\,.
\ee
The model approaches a free field theory for $g\rightarrow 0$.
To calculate the partition function  in the presence of sources
$J^a(x)$ it is convenient to
introduce the Lagrange multiplier field $G(x)$ as

\be
\label{ZJ}
Z[J^a] = \int ({\cal D} \Phi^a)({\cal D} G) e^{\textstyle
  -\frac{1}{2}\int \Phi^a (-\partial^2)\Phi^a +\frac{i}{2}\int
  G(\Phi^a\Phi^a - \frac{1}{g})+ \int J^a\Phi_a}\,,
\ee

\noindent
From (\ref{ZJ}) we see that is is natural to consider the effective
coupling $\hat{g}=gN$ which for large-$N$ may be adjusted to remain  $O(1)$ as
$g\rightarrow 0$. Integrating out the $\Phi^as$ and setting
$G(x)=G_0+\lambda(x)/\sqrt{N}$  we obtain the (renormalizable) $1/N$
expansion as

\be
\label{ZJ1}
\frac{Z[J^a]}{Z_0}= \int ({\cal D}\lambda) e^{\textstyle
  -\frac{N}{2}[Tr(\ln(1-\frac{i}{\sqrt{N}}\frac{\lambda}{-\partial^2})) +
    \frac{i}{\sqrt{N}} \frac{\lambda}{\hat{g}} ]}
e^{\textstyle \frac{1}{2}\int J^a
  \frac{1}{-\partial^2}(1
  -\frac{i}{\sqrt{N}}\frac{\lambda}{-\partial^2})^{-1}J^a}\;,
\ee

\noindent 
where $Z_0$ depends on $G_0$. The critical theory is obtained for
$G_0=0$ and the critical coupling is determined by the
gap equation
\be
\frac{1}{\hat{g}_*} = \int \frac{d^3 p}{(2\pi)^3} \frac{1}{p^2}\,.
\ee
The resulting generating functional takes the form

\be
\ba{l}

\dps \frac{Z_*[J^a]}{Z_0}=\int ({\cal D}\lambda) e^{\textstyle
  -\frac{1}{2}\int \lambda K \lambda
-\frac{i}{3!\sqrt{N}} \int\;
\begin{picture}(0,0)

  \put(7,-2){\includegraphics{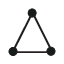}}

  \put(-3.7,-5.5){$\lambda$}
  \put(25,-5.5){$\lambda$}
  \put(11,15){$\lambda$}

\end{picture}
\qquad \quad -\frac{1}{8N}\int\;
\begin{picture}(0,0)

  \put(7,-2){\includegraphics{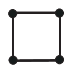}}

  \put(-3.7,-6){$\lambda$}
  \put(27,-6){$\lambda$}
  \put(27,14.8){$\lambda$}
  \put(-3.7,14.8){$\lambda$}

\end{picture}
\qquad\quad + \,\cdots
}
\\
\\

\hspace{1cm}\times \;\;
\dps e^{\textstyle
  \frac{1}{2}\int J^a \Big[
\begin{picture}(0,)

   \put(10,3){\line(1,0){18}}

\end{picture}
\qquad \quad + \frac{i}{\sqrt{N}}
\begin{picture}(0,0)

   \put(10,3){\line(1,0){18}}
   \put(19,2.9){\circle*{3}}
   \put(16,-7){\small $\lambda$}

\end{picture}
\qquad \quad - \frac{1}{N}
\begin{picture}(0,0)

   \put(10,3){\line(1,0){36}}
   \put(21,2.9){\circle*{3}}
   \put(35,2.9){\circle*{3}}

   \put(18,-7){\small $\lambda$}
   \put(32,-7){\small $\lambda$}
   \put(55,0){$+ \,\cdots \Big]J^a\;\;.$}

\end{picture}
\qquad\qquad}
\ea
\ee

The basic propagator of the $\Phi^a(x)$s is
\be
\Delta(x) = \frac{1}{4\pi}\frac{1}{(x^2)^{1/2}} \, \,\,\,\sim\,\,\,
\begin{picture}(0,0)

   \put(10,3){\line(1,0){40}}

\end{picture}
\ee
The operator $K$ and its inverse are then found to be
\be
K=\frac{\Delta^2}{2}\,,\,\,\,\,\,K^{-1} =
-\frac{16}{\pi^2}\frac{1}{x^4} \,\,\,\sim\,\,\, -
\begin{picture}(0,0)

   \put(5,-1){\includegraphics{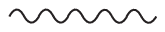}}

\end{picture}
\ee
Now we can calculate all n-pt functions of $\Phi^a$. For example, the
2-pt function is given by
\bea
\label{2ptf}
\langle \Phi^a \Phi^b \rangle  &=& \delta^{ab}\Bigl[
    \begin{picture}(0,0)

   \put(10,3){\line(1,0){60}}

\end{picture}
\qquad \qquad\quad\qquad -\frac{1}{N}
\begin{picture}(0,0)

\put(5,0){\includegraphics{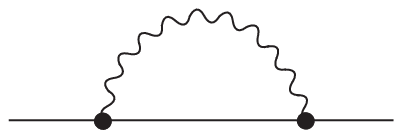}}

\end{picture}
\qquad \qquad\qquad \qquad \qquad 
+\cdots \Bigl] \nonumber \\
&=&\delta^{ab} \frac{1}{(x^2)^{1/2}} [1- \eta_1 \ln x^2 + ...]\;,
\qquad \eta_1 = \frac{1}{N}\frac{4}{3\pi^2}\;.
\eea
We notice that the elementary fields have acquired an anomalous
    dimension $\eta_1$. Ideally, a holographic description of the
    $O(N)$ vector model should reproduce this result from a bulk
    calculation, however, such a calculation is still elusive. On the
    other hand, bulk
    fields would give the correlation functions of composite
    boundary operators. The generating functional for one such 
    operator
    may be obtained if we
    consider a external source for the fluctuations of the 
auxiliary field as

\bea
&&\frac{Z[J^a]}{Z_0}\rightarrow Z[A]\nonumber \\
&&=\int ({\cal D}\lambda) e^{\textstyle -\frac{1}{2}\int \lambda K \lambda
-\frac{i}{3!\sqrt{N}} \int
\begin{picture}(0,0)

  \put(7,-2){\includegraphics{trianglesmall.eps}}

  \put(-3.7,-5.5){$\lambda$}
  \put(25,-5.5){$\lambda$}
  \put(11,15){$\lambda$}

\end{picture}
\qquad \quad -\frac{1}{8N}\int\;
\begin{picture}(0,0)

  \put(7,-2){\includegraphics{quadrat.eps}}

  \put(-3.7,-6){$\lambda$}
  \put(27,-6){$\lambda$}
  \put(27,14.8){$\lambda$}
  \put(-3.7,14.8){$\lambda$}

\end{picture}
\qquad\quad + \cdots +\int A\lambda  }\;.
\eea

\noindent
This can be viewed as the generating functional $e^{\hat{W}[A]}$ for a
    conformal scalar operator $\lambda$ with
a dimension $\Delta = 2+O(1/N)$. To be precise, since
\be
\int ({\cal D}\lambda) e^{\textstyle -\frac{1}{2}\int \lambda K
    \lambda+\int A\lambda}
= e^{\textstyle \frac{1}{2}\int A\Pi\lambda}\,,
\ee
and $\Pi$ would give a non-positive 2-pt function, we should
    actually consider
\be
W[A] = \hat{W}[iA]\;.
\ee
The proposal of \cite{KP} may then be concretely presented as
\be
e^{W[A]} \equiv \int_{AdS_4} ({\cal D}\Phi) e^{-I_{HS}(\Phi)}\;,
\ee
where
\be
\label{scal_act}
I_{HS}(\Phi) = \frac{1}{2\kappa^2_4}\int d^4x
 \sqrt{g}\Big(\frac{1}{2}(\partial
 \Phi)^2+\frac{1}{2}m^2\Phi^2 +
... \Big)\,,
\ee
with $m^2 =- 2$ that  corresponds to conformally coupled scalar
 on AdS$_4$. We see that the problem in hand naturally asks for a
 "bottom-up" approach, namely, use  the full knowledge of the boundary
 effective action in order to calculate the bulk path integral.
 Indeed, in principle  we should be able to have control of the fully
 quantized bulk theory, since bulk quantum corrections would correspond
 to the $1/N$ corrections of a renormalizable  boundary theory. For
 the time being, however,  one can be content if knowledge
 of the boundary generating functional for composite operators can
 help her calculate the elusive classical bulk action for a HS gauge
 theory.

Let me  briefly describe now, how this  "lifting program" works \cite{tassos}.
A possible form of the bulk HS
action is
\bea
\label{ads4_act}
I_{HS}(\Phi) &=& \frac{1}{2\kappa^2_4}\int d^4x
\sqrt{g}\Big(\frac{1}{2}(\partial \Phi)^2-\Phi^2 +
\frac{g_3}{3!}\Phi^3 +\frac{g_4}{4!}\Phi^4\nonumber \\
&&+\sum_{s=2}^{\infty} {\cal G}_s h^{\{\mu_1 ... \mu_s\}} \Phi\,
\partial_{\{\mu_1} ...\partial_{\mu_s\}}\Phi+\cdots\Big)\;,
\eea
where ${\cal G}_s$ denote the couplings of the conformally coupled
scalar to the higher-spin gauge fields $h^{\{\mu_1,..,\mu_s\}}$ that
can be taken to be totally symmetric traceless and conserved tensors.
From (\ref{ads4_act}), by the standard holographic procedure involving
the evaluation of the on-shell bulk action with specified
boundary conditions, we obtain schematically:
\\

-- 2-pt function --

\be
\langle \lambda \lambda  \rangle \quad \sim \quad
\begin{picture}(0,0)
\put(20,3){\circle{28}}
\put(6,3){\line(28,0){28}}
\put(6,3){\circle*{2}}
\put(34,3){\circle*{2}}
\end{picture}
\ee

-- 3-pt function --

\be
\langle \lambda \lambda \lambda  \rangle  \quad \sim \quad g_3
\begin{picture}(0,0)
\put(20,3){\circle{28}}

\put(10,12.5){\circle*{2}}
\put(30,13){\circle*{2}}
\put(20,-11){\circle*{2}}
\put(20,3){\circle*{2}}

\put(10,12.5){\line(1,-1){10}}
\put(20,3){\line(1,1){10}}
\put(20,-11){\line(0,1){13}}

\end{picture}
\ee
\\

The 4-pt function depends on $g_4$, $g_3^2$ and on {\it all} the
higher-spin couplings ${\cal G}_s$. It is technically not impossible
to write down the bulk tree graphs that involve the exchange of
higher-spin currents \cite{Ruhl}. Moreover, bulk gauge invariance
should, in principle, be  sufficient to determine {\it all} the ${\cal
  G}_s$
in terms of only one of them i.e. in terms of the coupling ${\cal
  G}_2$ of the energy momentum tensor.  Schematically we have

\be
\ba{c}
\dps
\langle \lambda \lambda \lambda  \lambda \rangle  \quad \sim \quad

g_4
\begin{picture}(0,0)
\put(20,3){\circle{28}}

\put(10,13){\circle*{2}}
\put(29.5,-7){\circle*{2}}

\put(30,12.7){\circle*{2}}
\put(10,-7){\circle*{2}}
\put(20,3){\circle*{2}}

\put(10.5,12.4){\line(1,-1){19}}
\put(10,-7){\line(1,1){20}}

\end{picture}

\qquad\qquad +\quad g_3^2\Biggr[
\begin{picture}(0,0)
\put(20,3){\circle{28}}

\put(10,13){\circle*{2}}
\put(29.5,-7){\circle*{2}}

\put(30,12.7){\circle*{2}}
\put(10,-7){\circle*{2}}

\put(10,13){\line(1,-1){20}}
\put(10,-7){\line(1,1){20}}

\end{picture}
\qquad\quad + {\rm crossed}

\Biggr]+\sum_s{\cal G}_s\Biggl[\cdots\Biggl]

\ea
\ee

The above must be compared with the corresponding result obtained from
the $W[A]$.
This would allow us to fix the scalings of various coefficients as (we set
the $AdS$ radius to 1)
\be
\frac{1}{2\kappa_4^2} \sim N\;, \quad g_3\,, g_4 \sim O(1)\;, \quad
{\cal G}^2_s\sim O(\frac{1}{N})\;\; {\rm iff}\;\; \langle h_s \;h_s\rangle
\sim O(1)\;.
\ee

The first result obtained this way was  \cite{tassos}
\be
 g_3 = 0\;.
\ee
This has been confirmed by a direct calculation using the Vasiliev
equations on AdS$_4$ \cite{SS}. Further results have been reported in
\cite{Ruhl}, however much more needs to be done to get a satisfactory
understanding of the bulk HS action. Another open issue is to
reproduce, from bulk loops, the known boundary anomalous dimension of
the operator $\lambda$. Finally, it remains to be understood
\footnote{See \cite{Ruhl_recent} for a recent
  work on that issue.}  how the
Higgs mechanism for the bulk HS fields gives rise to the known $1/N$
corrections of the anomalous dimensions of the corresponding boundary
higher-spin operators.

\section{The role of boundary double-trace transformations and the first
  trace  of duality}

In our study of the critical $O(N)$ vector model we have started with  an
elementary field $\Phi^a$ with dimension $\Delta=1/2+O(1/N)$ and
obtained
a composite operator $\lambda$ with $\Delta_\lambda = 2+O(1/N)$.
It follows that we are dealing with an interacting CFT even for
$N\rightarrow \infty$, since the free CFT would have had a composite
operator like
$\frac{1}{\sqrt{N}} \Phi^a\Phi^a$ with $\Delta_{\Phi^2}=1$. It is then
natural to ask
where is the free theory? To find it we consider the Legendre
transform of $W[A]$ as
\bea
W[A] +\int AQ &=& \Gamma[Q]\;,\\
\Gamma[Q] &=& \Gamma_0[Q]+\frac{1}{N}\Gamma_1[Q]+...\,,\\
\Gamma_0[Q] &=& \frac{1}{2}\int Q\,K\, Q+ \frac{1}{3!\sqrt{N}}\int\;
\begin{picture}(0,0)

  \put(7,-5){\includegraphics{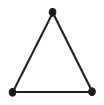}}

  \put(-5,-8){$Q$}
  \put(35,-8){$Q$}
  \put(15,26){$Q$}

\end{picture}
\qquad\qquad + \cdots\,.
\eea
The generating functional  for the correlation functions of the free field
$\frac{1}{\sqrt{N}} \Phi^a\Phi^a$ with dimension $\Delta=1$ is $\Gamma_0[Q]$.
Therefore, we face here the holographic description of a free
field theory!

The theory described by $\Gamma[Q]$ has imaginary couplings and
anomalous dimensions below the unitarity bounds
\cite{tassos_old}. Nevertheless, {\it both} the theories described by
$W[A]$ and
$\Gamma[Q]$ should be the holographic duals of a {\it unique} HS bulk
theory.  Moreover, it turns out that these two theories are related to
each other by an underlying dynamics that
appears to be generic in non-trivial models of three-dimensional
CFT. We propose
that this particular underlying dynamics is a kind of duality.
This is motivated by the fact that for $N\rightarrow\infty$ the
spectrum  of free field theory and 
the spectrum of theory $W[A]$ are {\it almost}
the same. Indeed, the
$W[A]$ theory contains conserved currents for
$N\rightarrow\infty$. The {\it only} difference between the two
theories at leading-$N$ is the interchange of the two scalar
operators corresponding to the two following UIRs of SO(3,2)
\be
\label{duality}
D(1,0) \longleftrightarrow D(2,0)\,.
\ee
These UIRs are {\it equivalent} as they are related by Weyl
reflections and have the same Casimirs.

The {\it duality} (\ref{duality}) is induced by a particular type of
dynamics usually
referred to as {\it double-trace
deformations}.\footnote{We keep this terminology despite the fact that there
are no traces taken here.} To show the essence of our proposal, consider
an operator $Q(x)$ with
a dimension $\Delta=1$ i.e. an operator in free field theory.
Then $Q^2(x)$ is a relevant deformation of a
theory and we can consider the deformed
2-pt function as
\bea
\label{QQ}
&&\langle Q(x_1)Q(x_2) \; e^{\textstyle \frac{f}{2}\int Q^2(x)}\rangle =
\langle Q(x_1)Q(x_2)  \rangle_f\nonumber \\
&&= \langle Q(x_1)Q(x_2)  \rangle_0 + \frac{f}{2}\int d^3x \langle
Q(x_1)Q(x_2)Q^2(x) \rangle_0 + ... \,.
\eea
We now make a large-$N$ factorization assumption such that, for
example,
\be
\frac{1}{2}\langle Q(x_1)Q(x_2) Q^2(x) \rangle_0 \simeq
\langle Q(x_1)Q(x)\rangle_0\langle Q(x_2)Q(x) \rangle_0 +
O\left(\frac{1}{N}\right)\,,
\ee
and similarly for all correlators that appear in (\ref{QQ}). We then
obtain
\bea
\langle Q(x_1)Q(x_2)  \rangle &=& \langle Q(x_1)Q(x_2)  \rangle_0 \nonumber \\
&&+ f\,\int d^3x \langle Q(x_1)Q(x)\rangle_0\langle Q(x_2)Q(x)
\rangle + O\left(\frac{1}{N}\right)\;.
\eea
In momentum space this looks like
\be
Q_f(p)= \frac{Q_0(p)}{1-fQ_0(p)}\,,\,\,\,\,\,Q_0(p)\simeq \frac{1}{p}\,.
\ee
In the infrared, i.e. for small momenta $|p|\ll f$, we find
\be
\label{Qexp}
f^2Q_f(p) = -\frac{f}{\dps  1-\frac{1}{fQ_0(p)}}\simeq -f-Q_0^{-1}(p) + ...\;.
\ee
If we drop the non-conformal constant $f$ term on the r.h.s. of
(\ref{Qexp}) we obtain the 2-pt function of an operator with dimension
$\Delta_f=2$. We see that the UV dimension $\Delta_0=1$ has changed to
the IR dimension  $\Delta_f=2$ and that this change is induced by  the
double-trace deformation.

The above  dynamics must be seen in $AdS_4$.
The on-shell bulk action of a conformally coupled scalar, using the standard
Poincar\'e coordinates,  is
\be
I_\varepsilon = -\frac{1}{2}\frac{1}{\varepsilon^2}\int d^3 x\,
\Phi(\bar{x};\varepsilon)
\partial_r \Phi(\bar{x};r)\Big|_{r=\varepsilon\ll 1}\;.
\ee
To evaluate it we need to solve the Dirichlet problem
\be
(\nabla^2-2)\Phi(\bar{x};r) = 0\;,\quad \Phi(\bar{x};r=\infty)=0\;,\quad
\Phi(\bar{x};\varepsilon) =  \Phi(\bar{x};\varepsilon)\;.
\ee
\be
 \Phi(\bar{x};r) = \int\frac{d^3 p }{(2\pi)^3}
 e^{i\bar{x}\bar{p}}\Phi(\bar{p};\varepsilon)\,
\frac{r}{\varepsilon}\, e^{-|p|(r-\varepsilon)}\;.
\ee
One way to proceed is via the Dirichlet-to-Neumann map \cite{Uhl}
that relates the boundary value of a field in a certain manifold
${\cal M}$ to its normal derivative at the boundary, e.g.
\be
\Phi(x)\Big|_{x\in\partial{\cal M}} = f(\bar{x})\;;
\quad \hat{\Lambda}f= n^{\mu}\partial_\mu\Phi(x)\Big|_{x\in\partial{\cal M}}\;,
\ee
where $n^\mu$ is the normal to the boundary vector. 
Knowledge of the map $\hat{\Lambda}$ allows the reconstruction of
the bulk metric. For the conformally coupled scalar
we have the remarkably simple expression
\be
\partial_r\Phi(\bar{p};r)\Big|_{r=\varepsilon} =
(\frac{1}{\varepsilon} - |p|)\Phi(\bar{p};\varepsilon)\;.
\ee
The terms in parenthesis on the r.h.s. may be viewed as a {\it
  generalized Dirichlet-to-Neumann map} since we have taken the
boundary to be at $r=\epsilon$. This map may be identified
with the r.h.s. of the expansion (\ref{Qexp}) is we set
$f=1/\epsilon$, such that
\be
\frac{1}{\epsilon^2}\left[\hat{\Lambda}_{\epsilon}(p)\right]^{-1} \sim
f^2 Q_f(p) =f^2\frac{Q_0(p)}{1-fQ_0(p)}\,. 
\ee
We then see that the inversion of the generalized
Dirichlet-to-Neumann map for  a conformally coupled scalar corresponds
to the resummation induced by a double-trace deformation on the
free boundary 2-pt function. Notice that the limit
$\epsilon\rightarrow 0$ drives the boundary theory in the IR.

Let us finally summarize in a pictorial way some of the salient
features of  the two types of
holography discussed above. In Fig.1 we sketch the standard
holographic picture for the
correspondence of IIB string-SUGRA/${\cal N}=4$ SYM. In Fig.2 we
sketch what we have learned so far regarding the holography of the HS
gauge theory on AdS$_4$.


\begin{figure}[!tbh]
\psfrag{lambda}{\tiny $0\,\,\,\,\,\,\,\,\lambda \,\,\,\,\,\,\infty$}
\psfrag{alpha}{\tiny $0\,\,\,\,\,\,\,\,\alpha '\,\,\,\,\,\,\infty$}
\centerline{\includegraphics[height=8.5cm]{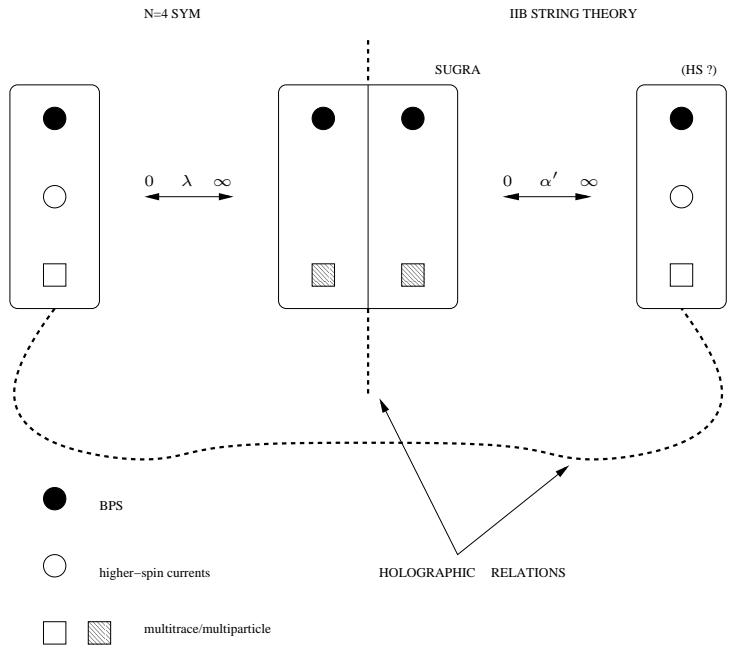}}

\vspace{0.2cm}
\caption{Type 1 holographic correspondence: ${\cal N}=4$ SYM/IIB
  string theory.}
\label{fig1}
\end{figure}

\begin{figure}[!tbh]
\centerline{\includegraphics[height=8.5cm]{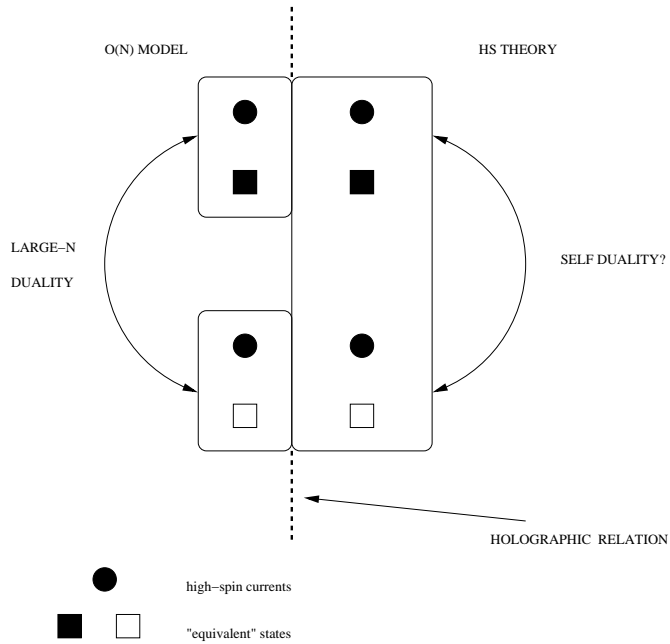}}
\vspace{0.2cm}
\caption{Type 2 holographic correspondence: O(N) vector model/HS on AdS$_4$.}
\label{fig2}
\end{figure}

\section{Double-trace deformations and the duality of linearized HS
 gauge theories}

In the previous section we emphasized the special role played by
the boundary double-trace deformation
\be
\label{dtr_scalar}
\frac{f}{2}\int d^3x\,Q(x) Q(x)\,,
\ee
for theories with bulk conformally invariant scalars.
This raises the possibility that boundary deformations of the form
\be
\label{dtr_HS}
\frac{f_s}{2}\int d^3x\,h^{(s)}h_{(s)}\,,
\ee
where $h^{(s)}$ denote symmetric traceless and conserved currents, may
play a crucial role in the holography  of HS gauge theories.
Such deformations are of course {\it irrelevant} for all $s\geq 1$,
nevertheless 
in many cases they lead to well-defined UV fixed points. Important
examples are the three-dimensional Gross-Neveu and Thirring models in
their large-$N$ limits
\cite{Anselmi}. We should emphasize that the large-$N$ limit is absolutely
crucial  for the well-defined and non-trivial nature of the above
results. 

For example, the dynamics underlying the Gross-Neveu model is,
in a sense, just the
opposite of the dynamics described in the previous section and is
again connected with the equivalence or ``duality'' between the irreps
$D(1,0)$ and $D(2,0)$ \cite{LP1}.
However, we face a problem with the idea that the deformations (\ref{dtr_HS})
for $s\geq 1$ induce the  exchange between equivalent irreps of
$SO(3,2)$. Starting with the irreps $D(s+1,s)$, $s\geq 1$, their
equivalent irreps are $D(2-s ,s)$. These, not only they fall below the
unitarity bound $\Delta\geq s+1$, but for $s>2$ they appear to
correspond to operators with negative scaling dimensions. Clearly, such
irreps cannot represent physical fields in a QFT. This is to be
contrasted with the case of the irreps $D(1,0)$ and $D(2,0)$ which
are both
above the unitarity bound.

The remedy of the above problem is suggested by old studies of
three- dimensional gauge theories. In particular, it is well-known
that to a
three-dimensional gauge potential $A_i(p)$ corresponds (we work in momentum
space for simplicity), a conserved
current $J_i(x) \propto i\epsilon_{ikj}p_jA_k(p)$, while to the
three-dimensional  spin-2 gauge potential $g_{ij}(p)$ corresponds a
conserved current (related to the Cotton tensor \cite{DJT}),
$T_{ij}(p) \propto \Pi^{(1.5)}_{ijkl}g_{kl}(p)$ (see below for the
definitions). A similar construction associates to each gauge field
belonging to the irrep $D(2-s,s)$ a physical
current in the irrep $D(s+1,s)$. Therefore, we need actually two steps to
understand the effect of the
double-trace deformations (\ref{dtr_HS}); firstly the deformation will
produce what it looks like a correlator for an operator
transforming under $D(2-s,s)$ that may be viewed as a gauge field,
secondly we transform this correlator
to one of a conserved current $D(s+1,s)$. Moreover, these
conserved currents
have opposite parity from the gauge fields, so we expect that parity
plays a role in our discussion.

Let us be more concrete and see what all the above mean in practice.
Consider a boundary CFT with a conserved current $J_i$ having momentum
space 2-pt
function
\be
\label{JJ0}
\langle {\cal J}_i {\cal J}_i \rangle_0 \equiv ({\cal J}_0)_{ij} =
\tau_1\frac{1}{|p|}\Pi_{ij}
+\tau_2\varepsilon_{ijk}p_k\,,\,\,\,\, \Pi_{ij}\equiv p_ip_j-\delta_{ij}p^2\,.
\ee
Consider the irrelevant double-trace deformation
\be
\frac{f_1}{2}\int {\cal J}_i {\cal J}_i\;,
\ee
and calculate
\be
({\cal J}_{f_1})_{ij} \equiv \langle {\cal J}_i {\cal J}_i
\,e^{\frac{f_1}{2}\int {\cal J}{\cal J}}\rangle
=({\cal J}_0)_{ij} +\frac{f_ 1}{2}\, \int \langle {\cal J}_i{\cal
  J}_j\Big|{\cal J}_k{\cal J}_k\rangle+ ...
\ee
Now assume: i) large-N expansion ${\cal J}_i{\cal J}_j \sim ({\cal
  J})_{ij} + O(1/N)$ and ii) existence of a UV fixed-point. The
leading-$N$ resummation yields
\be
f_1^2({\cal J}_{f_1})_{ij}  = \hat{\tau}_1 \frac{1}{|p|} \Pi_{ij}
+\hat{\tau}_2 \varepsilon_{ijk}p_k
\ee
\be
\hat{\tau}_1  \simeq \frac{f_1}{|p|}+\frac{1}{|p|^2}\frac{\tau_1}{\tau_1^2+\tau_2^2} + ...
\ee
\be
\hat{\tau}_2  \simeq-\frac{1}{|p|^2}\frac{\tau_2}{\tau_1^2+\tau_2^2} + ...
\ee
Dropping the  non-conformally invariant term $f_1/|p|$ we get
\be
f_1^2({\cal J}_{f_1})_{ij}  =
\frac{\tau_1}{\tau_1^2+\tau_2^2}\frac{1}{|p|^3} \Pi_{ij}
-\frac{\tau_2}{\tau_1^2+\tau_2^2} \varepsilon_{ijk}p_k\;.
\ee
This is the 2-pt function of a conformal operator $\hat{A}_i(\bar{p})$
transforming in the irrep  $D(1,1)$.
It lies below the unitary bound $\Delta\geq s+1$ of $SO(3,2)$,
therefore it must be a gauge field.
Define then the current $\hat{{\cal J}}_i =
i\varepsilon_{ijk}p_j\hat{A}_k$ that has 2-pt function
\be
\label{JJ1}
\langle \hat{{\cal J}}_i \hat{{\cal J}}_j\rangle  =
\frac{\tau_1}{\tau_1^2+\tau_2^2}\frac{1}{|p|^3} \Pi_{ij}
-\frac{\tau_2}{\tau_1^2+\tau_2^2} \varepsilon_{ijk}p_k\;.
\ee
It follows that there exists a "dual" theory with current $\hat{{\cal
    J}}_i$ that has 2-pt
function obtained from (\ref{JJ0}) by
\be
\label{tauS1}
\tau \rightarrow - \frac{1}{\tau}\,,\,\,\,\,\,\tau =\tau_2+i\tau_1\,.
\ee

Similarly, we may consider a boundary CFT having an energy momentum
tensor with 2-pt function
\be\label{TT}
\langle T_{ij}T_{kl}\rangle =\kappa_1
\frac{1}{|p|}\Pi_{ij,kl}^{(2)} - \kappa_2
\Pi_{ij,kl}^{(1.5)}\,,
\ee
where
\be
\ba{l}
\dps
\Pi_{ij,kl}^{(2)} = \half \Big[\Pi_{ik}\Pi_{jl} +\Pi_{il}\Pi_{jk}
  -\Pi_{ij}\Pi_{kl}\Big]\;,
\\
\\
\dps
\Pi_{ij,kl}^{(1.5)}=\frac{1}{4}\Big[
  \varepsilon_{ikp}\Pi_{jl}+\varepsilon_{jkp}\Pi_{il}
+\varepsilon_{ilp}\Pi_{jk}+\varepsilon_{jlp}\Pi_{ik} \Big]\;.
\\

\ea
\ee
The boundary irrelevant "double-trace"
deformation
\be
\frac{f_2}{2}\int T_{ij}T_{ij}\,,
\ee
leads (under the same large-$N$, existence of UV fixed point
assumptions as above), to a theory with an energy momentum tensor that
has 2-pt function obtained from (\ref{TT}) by \cite{LP2}
\be
\label{tauS2}
\kappa
\rightarrow -\frac{1}{\kappa}\;,\,\,\,\kappa=\kappa_2 +i\kappa_1
\,.
\ee
It is not difficult to imagine that the picture above generalizes to
all higher-spin currents in a three-dimensional CFT.

Now let us discuss what all the above boundary properties mean for the
bulk HS gauge theory. The first thing to notice is of course that the form
of the transformations (\ref{tauS1}) and (\ref{tauS2}) is reminiscent
of $S$-duality transformations. Then, we must ask what could be the
bulk action that yields (\ref{JJ0}) and (\ref{TT}). For example,
(\ref{JJ0}) may be the on-shell boundary value of the $U(1)$ 
action with a $\theta$-term on AdS$_4$
\be
\label{bulkEM}
I = \frac{1}{8\pi} \int d^4
x\sqrt{g}(\frac{4\pi}{e^2}F_{\mu\nu}F^{\mu\nu}+i
\frac{\theta}{2\pi}\frac{1}{2}
\varepsilon^{\mu\nu\rho\sigma}F_{\mu\nu}F_{\rho\sigma}
)\;.
\ee
We will show now that the bulk dual of the double-trace boundary
deformation discussed previously is a canonical duality transformation
i.e. a canonical
transformation that interchanges the bulk canonical
variables. We use the ADM form of  the Euclidean AdS$_4$ metric
(with radius set to 1)
\be
ds^2 = d\rho^2 +\gamma_{ij}dx^i dx^j\,,\,\,\,\,\,
\gamma_{ij}=e^{2\rho}\eta_{ij}\,,
\,\,\,\gamma=\mbox{det}\gamma_{ij}\,, \,\,\,
i,j=1,2,3\,.
\ee
to write the bulk action (\ref{bulkEM}) in terms of the canonical
variables as
\bea
\label{canEM}
I&=& \int d\rho \,d^3x \,\sqrt{\gamma}\left[\Pi^i\dot{A}_i-{\cal
    H}(\Pi^i,A_i)\right]\,, \\
{\cal H}(\Pi^1,A_i) &=& \frac{1}{e^2}
    \gamma^{-1}\gamma_{ij}\left({\cal E}^i{\cal E}^j -{\cal B}^i{\cal
    B}^j\right)\,, \\
\sqrt{\gamma}\Pi^i &=& \frac{2}{e^2}{\cal E}^i
    +i\frac{\theta}{4\pi^2}{\cal B}^i\,,\,\,\, {\cal E}^i
    =\sqrt{\gamma}E^i\,, \, {\cal B}^i =\sqrt{g}B^i\,,
\eea
with $E^i=F^{0i}$ and $B^i=\frac{1}{2}\epsilon^{ijk}F_{jk}$ the usual
    electric and magnetic  fields. The essence of the
    Hamilton-Jacobi approach to AdS holography is that for a given
    $\rho_0$  the variation of the
    bulk action with
    respect to the canonical variable $A_i$ gives, on shell,  the canonical
    momentum $\Pi^i$ at $\rho_0$. For $\rho_0\rightarrow \infty$ the
    latter is interpreted as the {\it 
    regularized} 1-pt function in the presence of
    sources, the reason being that requiring the regularity of the
    classical solutions inside AdS gives a relation between $\Pi^i$
    and $A_i$. Finally, to reach the boundary  one invokes
    a further technical step, (sometimes called holographic
    renormalization), such as to obtain finite 1-pt functions from which all
    correlation functions of the boundary CFT can be
    found. Schematically we have
\be
\label{HJ}
\frac{1}{\sqrt{\gamma}}
\left.\frac{\delta I}{\delta A_i(\rho_0,x_i)}\right|_{on-shell}
=\Pi^i(\rho_0,x_i) \sim_{\rho_0\rightarrow\infty} \langle {\cal
  J}^i(x_i)\rangle_{A_i}
\ee
In fact, one can show that for a $U(1)$ field  on AdS$_4$ there is no need for
renormalization as the solutions of the bulk e.o.m. give finite
contributions at the boundary.

Next we consider
canonical transformations in the bulk to move from the set variables
$(A_i,\Pi^i)$ to the new set $(\tilde{A}_i,\tilde{\Pi}^i)$. In
particular, we may consider
a generating functional of the 1st kind (see e.g. \cite{Goldstein}) of the form
\be
\label{F}
{\cal F}[A_i,\tilde{A}_i]=\frac{1}{2}
\int_{\rho =\,fixed}\!\!\!\!
d^3x\,\,\sqrt{\gamma}A_i(\rho,x_i)\epsilon^{ijk}\tilde{F}_{jk}(\rho,x_i)\,.
\ee
This induces the transformations
\bea
\label{PiB}
\frac{1}{\sqrt{\gamma}}\frac{\delta{\cal F}}{\delta A_i} &\equiv &\Pi^i
=\tilde{B}^i\,,\\
\label{BPi}
\frac{1}{\sqrt{\gamma}}
\frac{\delta{\cal F}}{\delta \tilde{A}_i} &\equiv &-\tilde{\Pi}^i
=B^i\,.
\eea
For $\theta=0$ these are  the standard duality transformations
$E^i\rightarrow B^i$, $B^i\rightarrow -E^i$.
We have at our disposal now two bulk actions, one written in terms of
$(A_i,\Pi^i)$ and the other in terms of $(\tilde{A}_i,\tilde{\Pi}^i)$,
which according to (\ref{HJ}) give at $\rho=\infty$,
\bea
\label{Ji}
\langle {\cal J}_i\rangle_{A_i} & = & \tilde{B}_i
=i\epsilon_{ijk}p_j\tilde{A}_k\,,\\
\label{tJi}
\langle \tilde{{\cal J}}_i\rangle_{\tilde{A}_i} & = & -B_i
=-i\epsilon_{ijk}p_jA_k\,.
\eea
From the above 1-pt functions we can calculate the corresponding 2-pt
functions by functionally
differentiating with respect to $A_i$ and 
$\tilde{A}_i$. We now take the following ansatz for the
the matrix $\delta \tilde{A}_i/\delta A_j$
\be
\label{AA}
\frac{\delta\tilde{A}_i}{\delta A_j} =C_1\frac{1}{p^2}\Pi_{ij} +C_2
  \epsilon_{ijk}\frac{p_k}{|p|} +(\xi-1)\frac{p_i p_j}{p^2}   \,,
\ee
where $\xi$ plays as usual the role of gauge fixing, necessary for its
  inversion. 
Then from (\ref{Ji}) and (\ref{tJi}) we find, independently of $\xi$  
\be
\label{JtJ}
\langle{\cal J}_i{\cal J}_k\rangle \langle \tilde{{\cal
    J}}_k\tilde{{\cal J}}_j
\rangle
=-\Pi_{ij}\,,
\ee
with $\Pi_{ij}$ defined in (\ref{JJ0}). 
It is then easy to verify that if $\langle{\cal J}_i{\cal J}_k\rangle$
    is given by the r.h.s. of (\ref{JJ0}), $\langle \tilde{{\cal
    J}}_k\tilde{{\cal J}}_j
\rangle$ is given by the r.h.s. of (\ref{JJ1}). In other words, we
    have shown that
    the bulk
    canonical transformation generated by (\ref{F}) induces the
    $S$-transformation (\ref{tauS1}) on the boundary 2-pt
    functions. We expect that our result generalizes to linearized bulk
    gravity in the Hamiltonian formalism\footnote{See \cite{HT} for a
    recent discussion of the duality in this context.} and linearized
    higher-spin theories.

An intriguing  property of the above transformations
generated by the boundary double-trace deformations 
is that combined with the  trivial transformation defined as
\be
\label{tau}
\tau \rightarrow \tau +1\;,
\ee
form the
SL(2,Z) group \cite{Witten}. The transformation (\ref{tau}) is the
boundary image of the 
bulk shift of the $\theta$-angle
\be
\theta \rightarrow \theta + 2\pi\,.
\ee
We expect that an analogous effect takes place in linearized
higher-spin gauge theories on AdS$_4$ when the appropriate
$\theta$-terms are introduced in the bulk
\cite{LP2,Vasiliev2}. The above suggest a special role for the SL(2,Z)
group in the study of HS gauge theories, even at the quantum level.

\section{Discussion}

It has been suggested \cite{Sundborg} that HS gauge theories emerge at
the tensionless limit of string theory, in a way similar to the
emergence of supergravity at the limit of infinite string tension. It
would be extremely interesting to quantify the above statement. A step
in this direction is the study of the holography of higher-spin
theories on AdS spaces. In this direction, both the study of specific
models, as well as investigations of generic holographic properties of
HS theories are important. The three-dimensional $O(N)$ vector model
provides a concrete example of a theory with a holographic HS dual. On
the other hand, bulk
dualities of linearized higher-spin theories may have far reaching
consequences for their CFT duals. 
For example, if a theory possesses a HS dual with a self-duality
property, its boundary double-trace
deformations despite being irrelevant might lead to a well-defined UV
completion of the theory. Also, it is conceivable that 
the self-duality property of linearized HS theories is a
remnant of a string theory duality in the tensionless
limit. Finally, it is well-known that 2-pt functions of
three-dimensional spin-1 
conserved currents can describe observable properties of Quantum Hall
systems \cite{BD}. It is then interesting to ask whether
boundary correlation functions of higher-spin currents may describe 
observables properties of physical systems. In particular, it
is intriguing to suggest \cite{LPnew} that linear gravity in AdS$_4$ may
correspond to special kinds of three-dimensional fluids in which
SL(2,Z) or a subgroup of it play a role.

\pagebreak

\Acknowledgements

{\it
I wish to thank the organizers of the First Solvay
  Workshop on ``Higher-Spin Gauge Theories'' for their invitation to
  present this work and for a well-organized and stimulating
  Workshop. Many thanks to Kostya Alkalaev for his help in preparing
  the manuscript.

The work presented here was done while the author
was in the University of Milano-Bicocca, supported by funds of the EU
Network HPRN-CT-2000-00131. I wish to thank the group at
Milano-Bicocca for their warm hospitality.}

\end{document}